\documentclass[]{spie}  
\usepackage[]{graphicx, epstopdf}
\usepackage{lscape}
\usepackage{amsmath}
\usepackage{wasysym}
\usepackage{afterpage}

\title{Orbital Differential Imaging: A New High-Contrast Post-Processing Technique For Direct Imaging of Exoplanets} 

\author{Jared R. Males\supit{a}, Ruslan Belikov\supit{b}, and Eduardo Bendek\supit{b}
\skiplinehalf
\supit{a}NASA Sagan Fellow, Steward Observatory, 933 North Cherry Avenue, Tucson, AZ, USA 85721 \\
\supit{b}NASA Ames Research Center, Moffett Field, CA 94035, USA
}

\authorinfo{Further author information: Send correspondence to J.R.M.: E-mail: jrmales@email.arizona.edu}

\begin{document} 
  \maketitle 

\begin{abstract}
Current post-processing techniques in high contrast imaging depend on some source of diversity between the exoplanet signal and the residual star light  at that location.  The two main techniques are angular differential imaging (ADI), which makes use of parallactic sky rotation to separate planet from star light, and spectral differential imaging (SDI), which makes use of differences in the spectrum of planet and star light and the wavelength dependence of the point spread function (PSF).  Here we introduce our technique for exploiting another source of diversity: orbital motion.  Given repeated observations of an exoplanetary system with sufficiently short orbital periods, the motion of the planets allows us to discriminate them from the PSF.  In addition to using powerful PSF subtraction algorithms, such an observing strategy enables temporal filtering.  Once an orbit is determined, the planet can be ``de-orbited'' to further increase the signal-to-noise ratio.  We call this collection of techniques Orbital Differential Imaging (ODI).  Here we present the motivation for this technique, present a noise model, and present results from simulations.  We believe ODI will be an enabling technique for imaging Earth-like planets in the habitable zones of Sun-like stars with dedicated space missions.
\end{abstract}



\section{INTRODUCTION}
\label{sec:intro}  
The challenge of imaging Earth-like exoplanets directly, that is resolving a planet's light from its host star's light, is typically stated along the lines of: ``achieve a $10^{-10}$ contrast ratio between planet and star at 0.1'' separation.''  This corresponds to a roughly 1 $R_e$ exoplanet orbiting 1 AU from a star 10 pc away.  The underlying technical challenges in coronagraphy and wavefront control are discussed at length in these proceedings and elsewhere.  

Here we are concerned with an approach to imaging Earth-like planets which to some extent relaxes the coronagraph and wavefront control requirements.  In short, it is possible that a space telescope dedicated only to observing the stars $\alpha$ Centauri A \& B for 2 years could image an Earth-like planet while only requiring $\sim$$10^{-8}$ raw contrast from the coronagraphic system. Given the proximity and brightness of these stars, only a $D$$\sim$45 cm telescope is required to well-resolve the habitable zones (HZs) of these stars.  For a complete introduction to this concept see the contributions by Bendek et al.\cite{bendek_spie15} and Belikov et al.\cite{belikov_spie15} in these proceedings.

One of the keys to this idea is the possibility that post-processing of the data will yield the additional factor of at least 100 in starlight suppression needed to image a planet.  Calibration and subtraction of the point spread function (PSF) is now a well developed art in ground-based high-contrast imaging.  The most widely used techniques are based on angular differential imaging\cite{2006ApJ...641..556M} (ADI).  In ADI the pupil is held in a fixed orientation with respect to the sky, usually by fixing the instrument rotator on an Alt-Az telescope.  Parallactic rotation will then cause the position of objects around the star to change with respect to the PSF.  An optimum PSF is calculated by a combination of the un-rotated images, using algorithms we discuss further below.  This optimum PSF is then subtracted from each image.  Any long-lived (``quasi-static'') speckles will be in the  PSF and so removed from the target image, acting as a high-pass filter (both spatially and temporally). The psf-subtracted images are then rotated to the same orientation (i.e. North up), and combined.  This then breaks the coherence of any remaining speckles, allowing for $\sqrt{N}$ reduction in the speckle noise.  Here we employ a similar concept.  Instead of parallactic rotation we will use the orbital motion of the planet to both provide diversity between images of the planet and measurement of the PSD, and to break the speckle coherence.  In addition, gathering a long time-series of data which (possibly) contains a signal modulated by well understood orbital mechanics lends itself to temporal (rather than spatial) analysis.

We first describe the basic steps of what we call ``Orbital Differential Imaging'' (ODI), given its heritage in ADI.  We then present initial results from a software simulator we are developing to prove this concept.

\section{The Steps of ODI}

ODI consists of three main steps: PSF subtraction, temporal filtering, and orbital co-adding. We now introduce each of these components.

\subsection{PSF Subtraction}

In order to remove long lived speckles we need to form an estimate of the PSF for each image.  As discussed above, in ground based work the most widely used technique is to allow parallactic rotation to move the planet with respect to the features of the PSF over the course of an observation.  Here we instead use the orbital motion of a planet, which allows us to choose images from the dataset which do not have planet signal in them at the same location as in the image being reduced.  

In ADI, the set of images nearest in time are usually excluded from the PSF.  The number of images excluded is typically set by the time if takes for an object to rotate by $\sim$1 FWHM so that there is little or no signal from the object in the PSF estimate.  Similarly, we can exclude images based on orbital motion.  In terms of the PSF full-width at half-maximum (FWHM) the speed of an orbiting planet projected onto the focal plane is given by 
\begin{equation}
v_{FOC} = 0.0834\left(\frac{D}{1\mbox{m}}\right) \left(\frac{1\mu\mbox{m}}{\lambda}\right) \left(\frac{1\mbox{pc}}{d}\right) \sqrt{ \left(\frac{M_*}{1M_{\astrosun}}\right)\left(\frac{1\mbox{AU}}{a}\right)} \mbox{ in FWHM day}^{-1}.
\label{eqn:vfoc}
\end{equation}
for a face-on circular (FOC) orbit \cite{2013ApJ...771...10M}.  Here $D$ is the telescope diameter, $\lambda$ is the observing wavelength, $d$ is the distance to the star, $M_*$ is the mass of the star, and $a$ is the semi-major axis.  See Males et al (2013) for the equations to project $v_{FOC}$ for an arbitrarily oriented orbit\cite{2013ApJ...771...10M}.  

Now if we observe for at least $1/v_{FOC}$ days, the planet will have moved 1 FWHM.  At a given location in a sequence of images longer than this characteristic time, there will be images with no planet signal.  These can then be used to determine a PSF.  So instead of angular differences, we  will make use of the orbital motion to perform differential imaging.  So images at times within
\[
t_{excl} = \delta/v_{FOC} \pm nP_{orb}
\]
where $n$ goes from 0 to the number of orbits of  period $P_{orb}$ contained in the sequence, should be excluded from the PSF.  The parameter $\delta$ specifies the number of FWHM to exclude.  It is nominally $\sim$1, and can be adjusted to optimize the reduction.

To determine the optimum PSF to subtract from each image, we use the Karhunen Lo'eve Image Processing (KLIP) algorithm \cite{2012ApJ...755L..28S}.  KLIP determines the optimum PSF in the least-squares sense by finding the Karhunen Loe've (KL) orthogonal basis set, or principle components\footnote{This algorithm is also commonly referred to as ``Principle Component Analysis'', or PCA}, of the set of reference images.  The PSF is determined by projecting the target image onto the KL basis.   In ODI we form the KL basis from the images not within $t_{excl}$ for the image being reduced.

\subsection{Temporal Filtering}
\label{sec:orbpsd}
The discussion of PSF subtraction above was primarily concerned with removing spatially correlated features, that is speckles, from the images.  Given a long time-series of images, it is natural to also attempt to analyze these data in the time and/or temporal-frequency domains.  Here we illustrate the potential for this analysis using the power spectral density (PSD) of an orbit.

Consider a single pixel at a projected separation of 1 AU from a 1 $M_{\astrosun}$ mass star.  If a planet is in a face-on circular orbit, it will pass through this pixel once per year.  We would then build up a time-series of measurements which include flux from that planet.  To illustrate, we assume a geometric albedo of 0.3, and use the Lambert phase function at quadrature, giving a contrast of $1.7\times10^{-10}$ for a 1 $R_{e}$ planet.  Finally, using an Airy pattern and the default observation parameters, we calculated such a time series and then calculated the power spectrum, $\mathcal{P}_{orb}$.

\begin{figure}[h]
\centering
\includegraphics[height=4in,clip=true,angle=90]{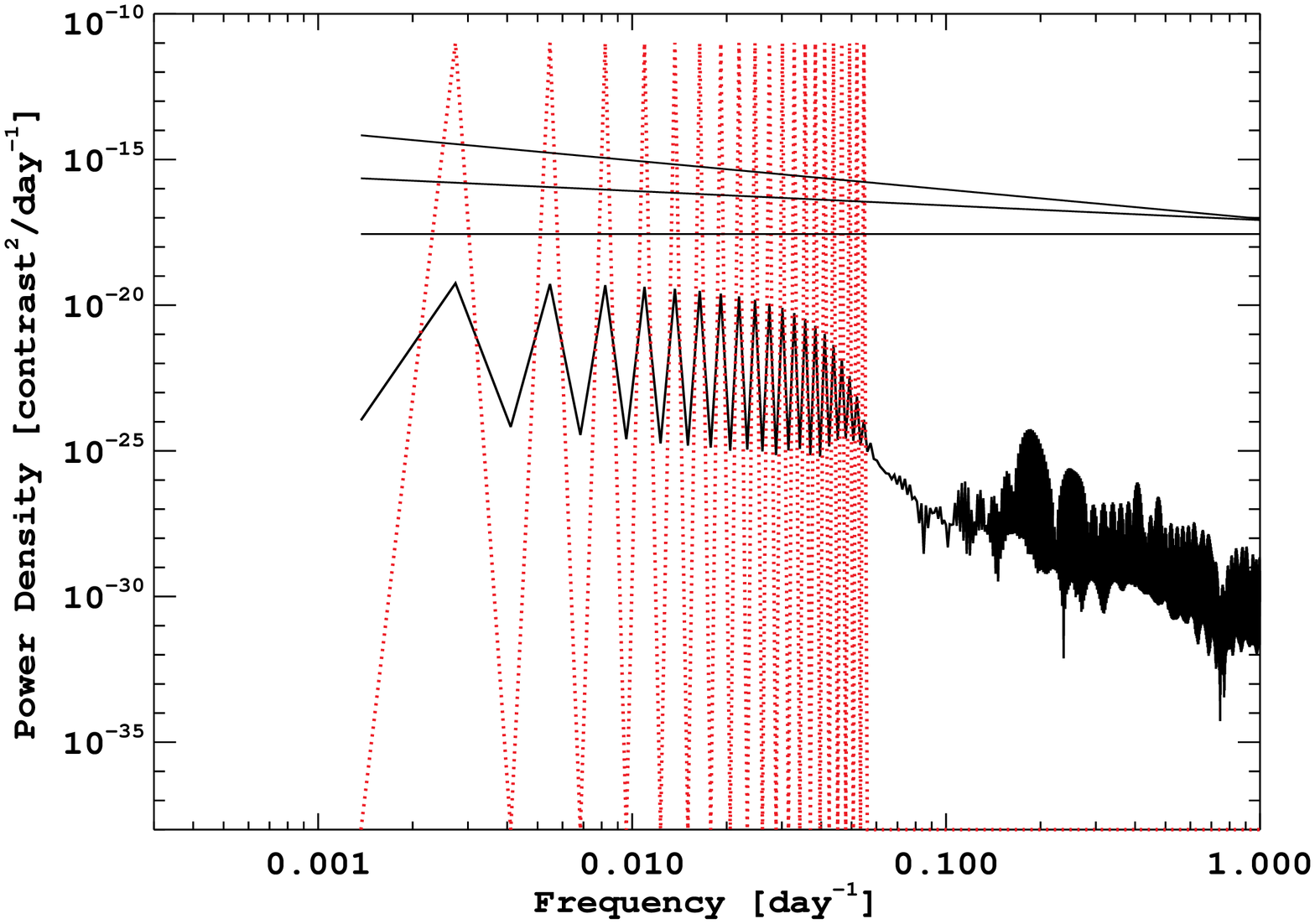}
\includegraphics[height=4in,clip=true,angle=90]{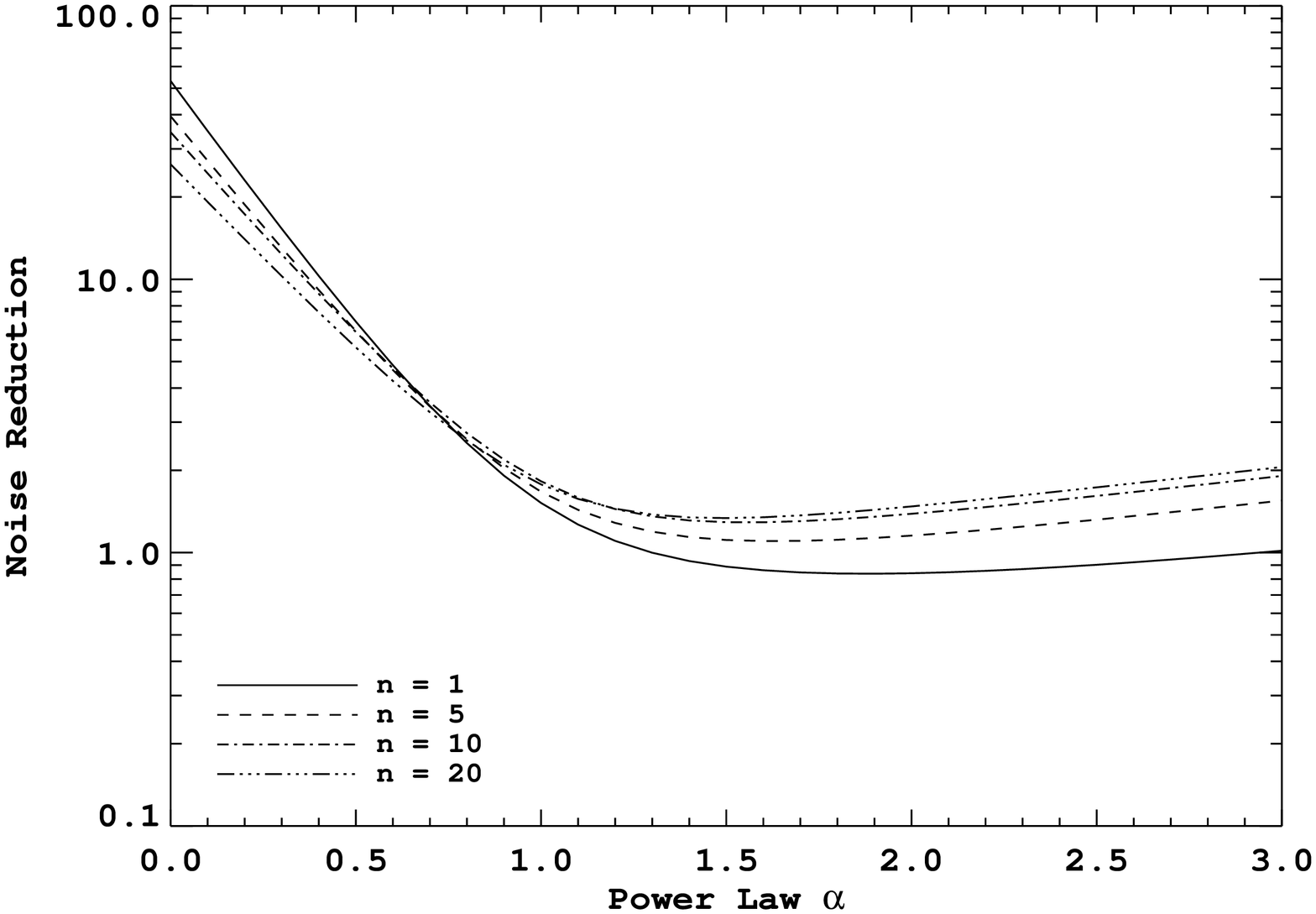}
\caption{Top: The PSD of the planet signal.  The main features are a series of narrow peaks at the orbital frequency, $1/P_{orb}$, and harmonics.  Also shown are speckle PSDs for $\alpha = 0,1,2$, normalized so that the speckle noise has $1\times10^{-8}$ contrast.  The red dotted line shows the simple comb filter described by Equation (\ref{eqn:comb_filter}) with $n=10$, scaled for display purposes. Bottom: the noise reduction due to applying the comb filter vs. the PSD power law $\alpha$.  Higher alpha implies more power at low temporal frequencies, so we see that temporal filtering is most effective for higher temporal frequency noise.  This has important implications for the ordering of the ODI steps. 
\label{fig:orbit_psd}}
\end{figure}

We present the orbital PSD in Figure \ref{fig:orbit_psd} (top panel).  We compare this to $1/f^{\alpha}$ noise PSDs with $\alpha=0,1,2$.  The noise PSDs are normalized so that their integrated power corresponds to $1\times 10^{-8}$ contrast.  The orbit PSD is characterized by a series of peaks, starting at $n/P_{orb}$.  These are the orbital frequency (0.0027 day$^{-1}$) and its harmonics.  This motivates use of a simple comb-filter, that is a series of band-passes at each harmonic, given by

\begin{equation}
   \Pi(f) = \left\{
     \begin{array}{lr}
       1, & 0-w < |f| < 0+w \\
       1, & \frac{i}{P_{orb} + w/i} < |f| < \frac{i}{P_{orb}-w/i} \mbox{ and } i \le n  \\
       0, & \mbox{otherwise}
     \end{array}
   \right.
   \label{eqn:comb_filter}
\end{equation} 
where $n$ specifies the maximum harmonic to use, and $w$ is the passband width. Such a filter is shown in red in Figure \ref{fig:orbit_psd}.  It is then straightforward to calculate the reduction in speckle variance\footnote{In this analysis we always present noise reduction as the ratio of raw noise to processed noise, such that numbers larger than 1 are an improvement} due to the filter:
\[
\frac{VAR_{sp}}{VAR_{sp,filt}} = \frac{\int\limits_{f_{min}}^{f_{max}} \mathcal{P}_{sp}(f) df} {\int\limits_{f_{min}}^{f_{max}} \mathcal{P}_{sp}(f) \Pi(f) df}.
\]
where $\mathcal{P}(f)$ denotes the temporal PSD. We must also account for the reduction in the peak signal from the planet, that is
\[
\frac{I_{orb, filt}^{peak}}{I_{orb}^{peak}} = \frac{\mbox{max}(\mathcal{F}^{-1}( \mathcal{F}(I_{orb}(t)) \Pi(f) ))}{\mbox{max}(I_{orb}(t))}
\]
where $\mathcal{F}$ is the Fourier transform.  The resultant reduction in noise is presented in the bottom panel of Figure \ref{fig:orbit_psd}.  We are applying a band-pass filter, which is preferentially passing low frequencies.  It is clear that this is most effective for higher temporal frequency noise, e.g. smaller $\alpha$.  This makes it complementary to PSF subtraction, which is a high-pass filter.

\afterpage{\clearpage}

\subsection{Orbital Co-adding}
\label{sec:saa}
A key step in ADI is the de-rotation of the images followed by combining or co-adding to beat down the residual speckle noise.  In ODI the analogous step is to shift each image according to the equations of orbital motion.  If we shift and add (SAA) along an orbit, we will co-add $\sim$$n P_{orb} v_{foc}$ independent images of the planet. Here $n$ is the number of orbital periods $P_{orb}$, and $v_{foc}$ is the orbital speed of the planet in the focal plane as before.   The reduction in the variance-of-the-mean will go as 
\[
{\sigma^2} = \frac{\sigma^2_o}{N_{sp}}
\]
where $\sigma^2_o$ is the variance in a single frame and $N_{sp}$ is the number of independent speckle realizations along the orbit.  There are 3 distinct regimes:
\begin{enumerate}
\item When speckles last for less than the characteristic orbital motion time, $1/v_{foc}$, then $N_{sp}$ is simply the number of speckle lifetimes during the integration.  There will be $1/(v_{foc} \tau_{sp})$ speckle realizations per image of the planet, so we have $N_{sp} = n P_{orb}/\tau_{sp}$.\\

\item For speckles which live longer than $1/v_{foc}$, but less than $P_{orb}$, the number of realizations per characteristic orbital motion time can be treated as 1, and so the total number of realizations is the number of $FWHM$s along the orbit.  That is $N_{sp} = nP_{orb}v_{foc}$.\\

\item In the case where speckles live longer than the orbital period, then we must account for the fact that the same speckle may be present on a subsequent orbit while we are co-adding.  So now the number of realizations is characterized by the number of speckles per orbit, $P_{orb}/\tau_{sp}$.
\end{enumerate}

\begin{figure}[h]
\centering
\includegraphics[height=4in,clip=true,angle=90]{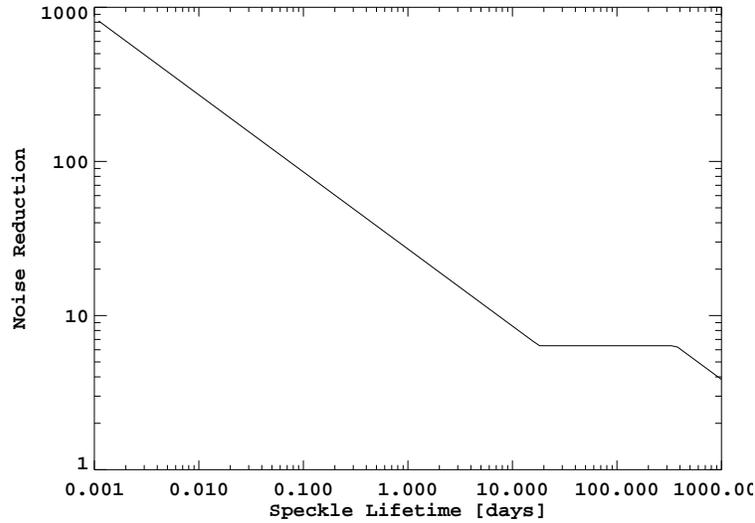}
\caption{Noise reduction due to the shift-and-add.  
\label{fig:speckle_saa}}
\end{figure}

\noindent We combine these regimes in Equation \ref{eqn:redsaa}, and present them graphically in Figure \ref{fig:speckle_saa}.  
\begin{equation}
   N_{sp} = \left\{
     \begin{array}{lr}
       \frac{n P_{orb}}{\tau_{sp}} &  \tau_{sp} < 1/v_{foc} \\
       nP_{orb}v_{foc}  &  1/v_{foc} \le \tau_{sp} < P_{orb} \\
       \frac{nP_{orb}^2}{\tau_{sp}}v_{foc} & P_{orb} \le \tau_{sp}
     \end{array}
   \right.
   \label{eqn:redsaa}
\end{equation} 

%

\section{Simulations}

The previous section describes the rational behind ODI, and lays out the 3 main steps.  Now we demonstrate the potential of ODI by analyzing a simplified simulation of a representative mission: ACESat\cite{bendek_spie15, belikov_spie15}.  ACESat is a proposed mission dedicated to searching for Earth-like planets in the HZs of the $\alpha$ Cen A \& B system. The hardware we simulated included the following:
\begin{enumerate}
\vspace{-5pt}
\itemsep0pt
\item 0.45m telescope primary, assumed ideally imaged onto the coronagraphic entrance pupil plane
PIAA coronagraph design similar to the Exo-C mission concept (Stapelfeld et al. 2015), except the radius of the occulter mask was 2 $\lambda/D$ (corresponding to about 1.8 $\lambda/D$ inner working angle)
\item 5 $\sim$$10\%$ bands with central wavelengths 420 nm, 468 nm, 520 nm, 580 nm, 646 nm. Each band was propagated monochromatically at its central wavelength for simplicity 
\item Science plane detector with photon noise dominating all other forms of detector noise (which were therefore neglected)
\end{enumerate}

\noindent Mission parameters were:
\begin{enumerate}
\vspace{-5pt}
\itemsep0pt
\item 2 year mission duration 
\item We assumed the mission cycled between the two stars and bands resulting in a 12 day cycle.  That is, 1 day per band per star and band, with a 1 day re-pointing gap.  This sequence was chosen primarily for convenience in simulator development.
\end{enumerate}

\noindent Astrophysical assumptions:
\begin{enumerate}
\vspace{-5pt}
\itemsep0pt
\item Stellar parameters for $\alpha$ Cen A from Bruntt et al. (2010)\cite{2010MNRAS.405.1907B}
\item Stellar input spectrum treated as a blackbody, scaled to match luminosity.  
\item No zodi or exozodi background (as long as this background is $< 10^{-8}$ contrast and temporally invariant, it would be subtracted by ODI)
\item Star system with 3 planets on circular orbits at $79^o$ inclination, which corresponds to the $\alpha$ Cen A-B binary orbit\cite{2002A&A...386..280P}.  The planet semi-major axes were: for Venus-like, 0.9 AU; for Earth-like, 1.2 AU; for pseudo-Mars, 2.5 AU.
\end{enumerate}
Note that the Venus-like and Earth-like orbits are scaled from the Solar system by $\sqrt{L_*}$, which we assume causes the photon budget at their atmospheres to be the same as in our Solar system.  This places the Earth-like planet in the habitable zone.  The pseudo-Mars planet parameters were selected to both fill out the FOV of the coronagraph, and be 50\% brighter than Mars. Albedo models were applied to each band to generate the reflected spectrum of the planet.

Given the above assumptions, the simulator constructs the speckle field I(x,y ; t) as follows. Wavefront control was abstracted by directly simulating a random post-wavefront control residual error (in the pupil plane) with properties expected from modern wavefront control methods. Specifically we assumed the following post-WC residuals in each image:
\begin{enumerate}
\vspace{-5pt}
\itemsep0pt
\item Random tip and tilt with 0.5 mas rms (each), uncorrelated between images (a white PSD). 
\item Constant high-order wavefront error in the pupil plane with a spatial $\alpha=3$ power law PSD. The level of this error was scaled to generate $10^{-8}$ speckle field in the image plane. This phase screen was modulated by the random tip/tilt error.
\end{enumerate}
Although simulating full end-to-end wavefront control is important, it is outside the scope of this paper.  Rather than making claims about achieved performance, the above list of assumptions can be treated as a requirements list for raw WFC residual. In particular, as we will show below, $10^{-8}$ raw contrast is sufficient for detection of $10^{-10}$ planets at SNR = 5 with ODI, even if 0.5 mas tip/tilt residual is added. This 0.5 mas can be relaxed by using a larger inner working angle, however, coronagraph designs continue to advance resulting in coronagraphs that are ever more tolerant to tip/tilt error without sacrificing performance.  

Finally, the star and planet plane waves were modulated by the above wavefront control residual in the coronagraphic entrance pupil plane (PIAA M1), separately propagated through to the science camera and then added together at the end, with photon noise, to simulate an exposure. A 2 year mission was simulated, changing filters and stars with the cadence described above, in 20 minute individual time steps.  In this case the observations of $\alpha$ Cen B were not simulated.

Once a 2 year mission simulation was completed, data reduction proceeded as follows.  The 20 minute exposures in each band were co-added into 1 day chunks.  Each band was then reduced using the ODI recipe:
\begin{enumerate}
\vspace{-5pt}
\itemsep0pt
\item The KLIP algorithm was applied using a well tested pipeline\cite{2014ApJ...786...32M} adapted for this purpose, assuming $1/v_{foc}$=19 days and $\delta$ (the exclusion parameter) varying from 0.5 to 1.5 depending on the band.
\item Temporal filtering was carried out by Gaussian smoothing each pixel in the time domain.  The more complicated Fourier domain filtering we described above was not implemented in this analysis.
\item For each 5-day/5-band set of observations, the images were transformed to the Red-Green-Blue (RGB) color system for ``true-color'' rendering.
\item Finally, each planet was reduced by shift-and-adding the images along its orbit.  The other planets were masked during this step, and the images were weighted by the normalized phase function of the planet.
\end{enumerate}

We show the results of this simulation in Figure \ref{fig:simulation}, which shows individual 5-day frames after KLIP and temporal filtering, but before de-orbiting.  In individual 5-day true-color frames both the Venus-like and the Earth-like planet are clearly detected.  

Unless one knows where to look, however, the pseudo-Mars planet is difficult to spot.  So we next performed the orbital SAA step.  The results are shown in Figure \ref{fig:simulation_saa}.  Note that in this case the orbital-SAA is easily performed as the pseudo-Mars planet signal is detectable at low S/N in individual frames.  An important area of investigation will be the performance of this technique when there is no prior information with which to determine the orbits.

\section{Conclusions}

We have presented an introduction to a new post-processing technique called orbital differential image, or ODI.  It is essentially a straightforward adaptation of the well-known ADI technique from ground-based imaging.  Where ADI uses parallactic rotation to differentiate speckles from planets, ODI instead uses orbital motion.  We can apply standard high-contrast imaging PSF subtraction algorithms, and given a long time baseline can employ temporal filtering.   

To begin testing ODI we have developed a mission simulator, which simulates propagation of wavefronts through a coronagraph, with injected aberrations.  With our initial assumptions, which included a $10^{-8}$ contrast raw residual and a 0.5 mas rms per-frame jitter, we were able to detect simulated Earth and Venus analogs using a full 2 years of data.  With the additional post-processing step of de-orbiting and co-adding, a much fainter pseudo-Mars was also detected.

These initial steps have demonstrated the feasibility of ODI.  We continue to refine the analytical justification for ODI and continue to develop the mission simulator.  Specific tasks currently underway include improvement of low-order aberration modeling and modeling of the time-evolution of high orders.  We are also beginning to test different operational strategies, for instance different dwell times in each pass band.  Further improvement in PSF subtraction is expected from using so-called reference differential imaging, where images from $\alpha$ Cen B are used to reduce those from $\alpha$ Cen B (and vice versa), and from using spectral differential image, where we will use the spectral evolution of the PSF.

\acknowledgments     
 
J.R.M was supported under contract with the California Institute of Technology (Caltech) funded by NASA through the Sagan Fellowship Program.  This work was supported in part by the National Aeronautics and Space Administration’s Ames Research Center.  Any opinions, findings, and conclusions or recommendations expressed in
this article are those of the authors and do not necessarily reflect the views of the National Aeronautics and
Space Administration.

\begin{landscape}
\begin{figure}
\centering
   \includegraphics[height=5cm]{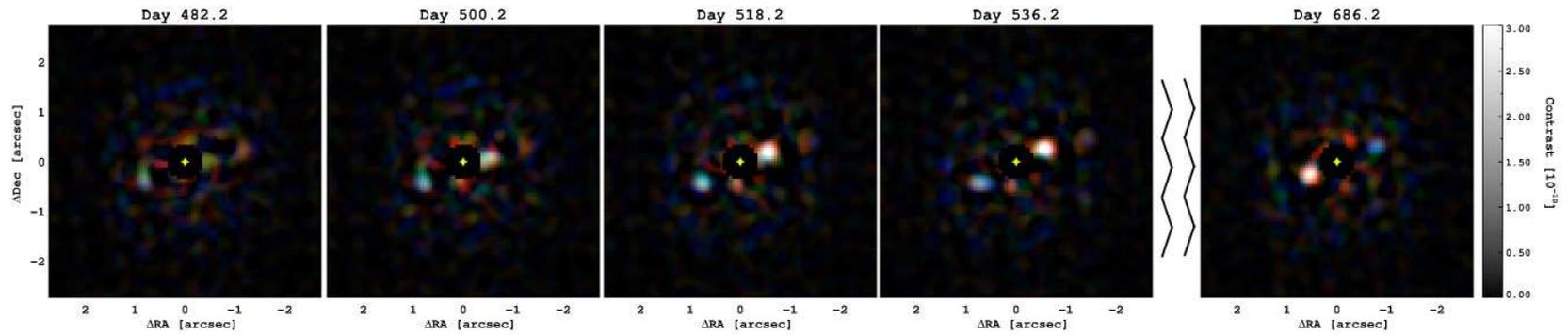}
   \caption[Simulation] 
   { A select sequence of images produced by ODI reductions of a simulated 2 year ACESat mission.  Observations in the 5 photometric bands were combined in 5 day chunks by transforming to the RGB color system.  The white/grey planet has a Venus-like spectrum, and the pale-blue dot is an Earth-like planet on a $\sqrt{L_*}$ orbit.  
     \label{fig:simulation} 
   }
\end{figure}

\begin{figure}
\centering
   \includegraphics[height=7cm,angle=90]{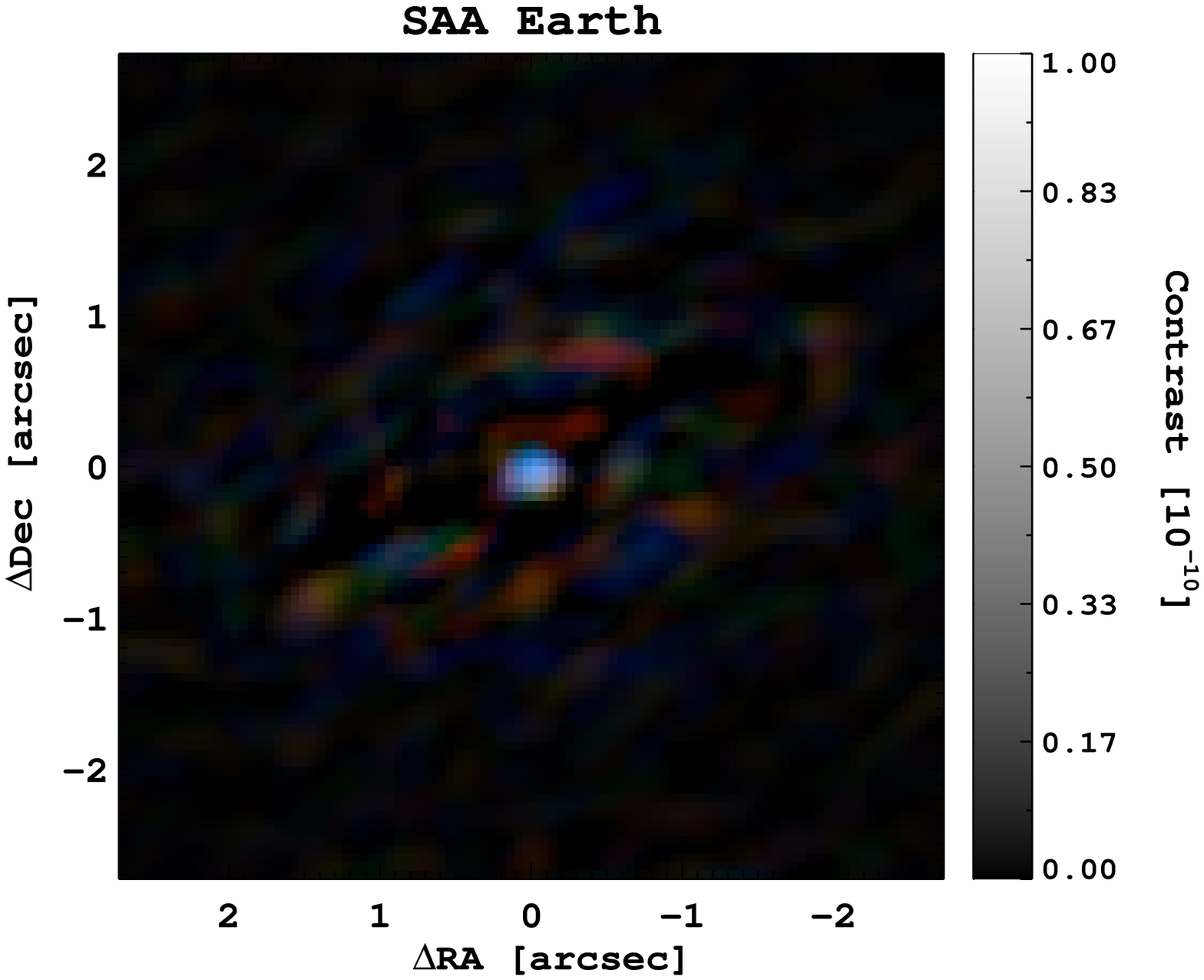}
   \includegraphics[height=7cm,angle=90]{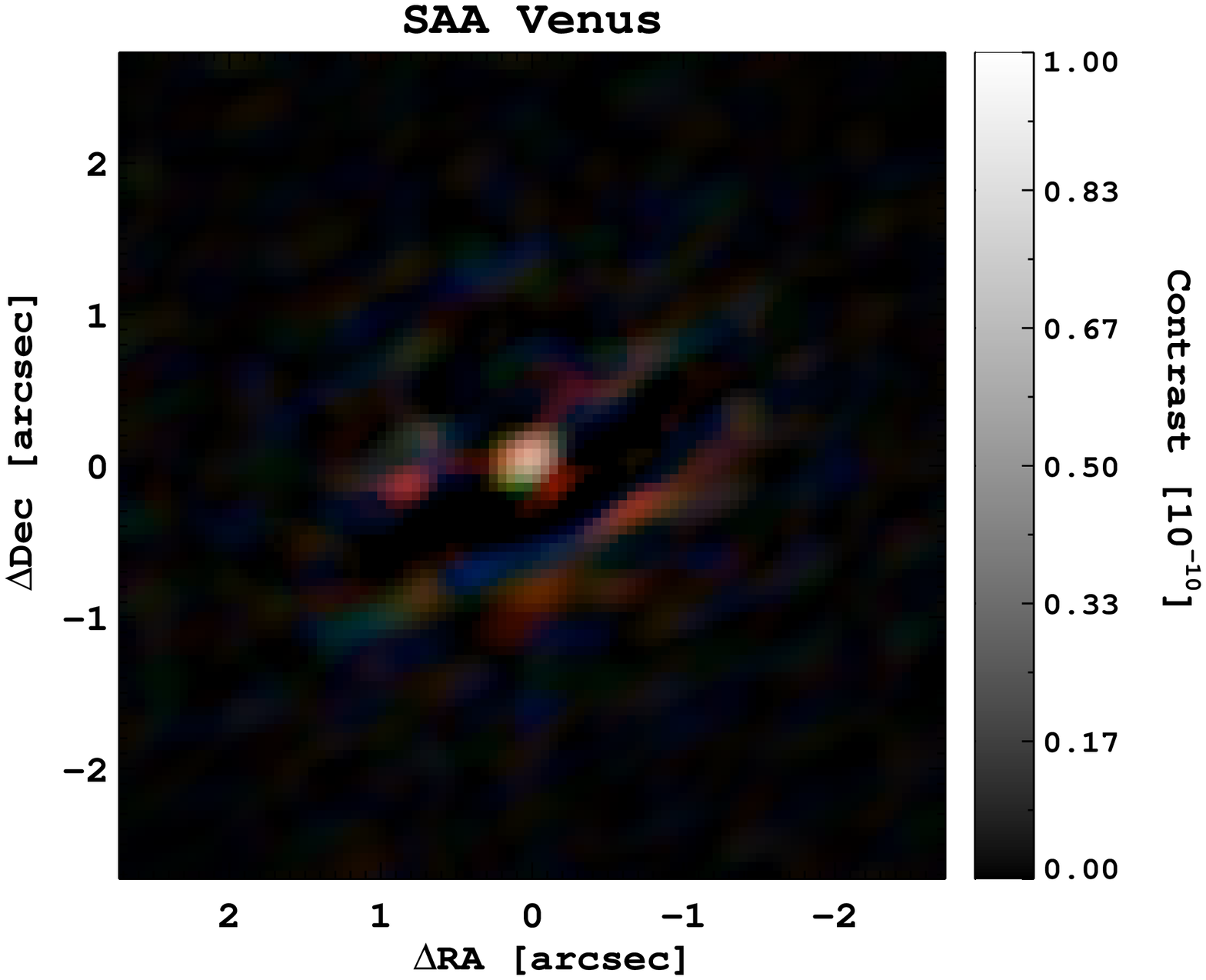}
   \includegraphics[height=7cm,angle=90]{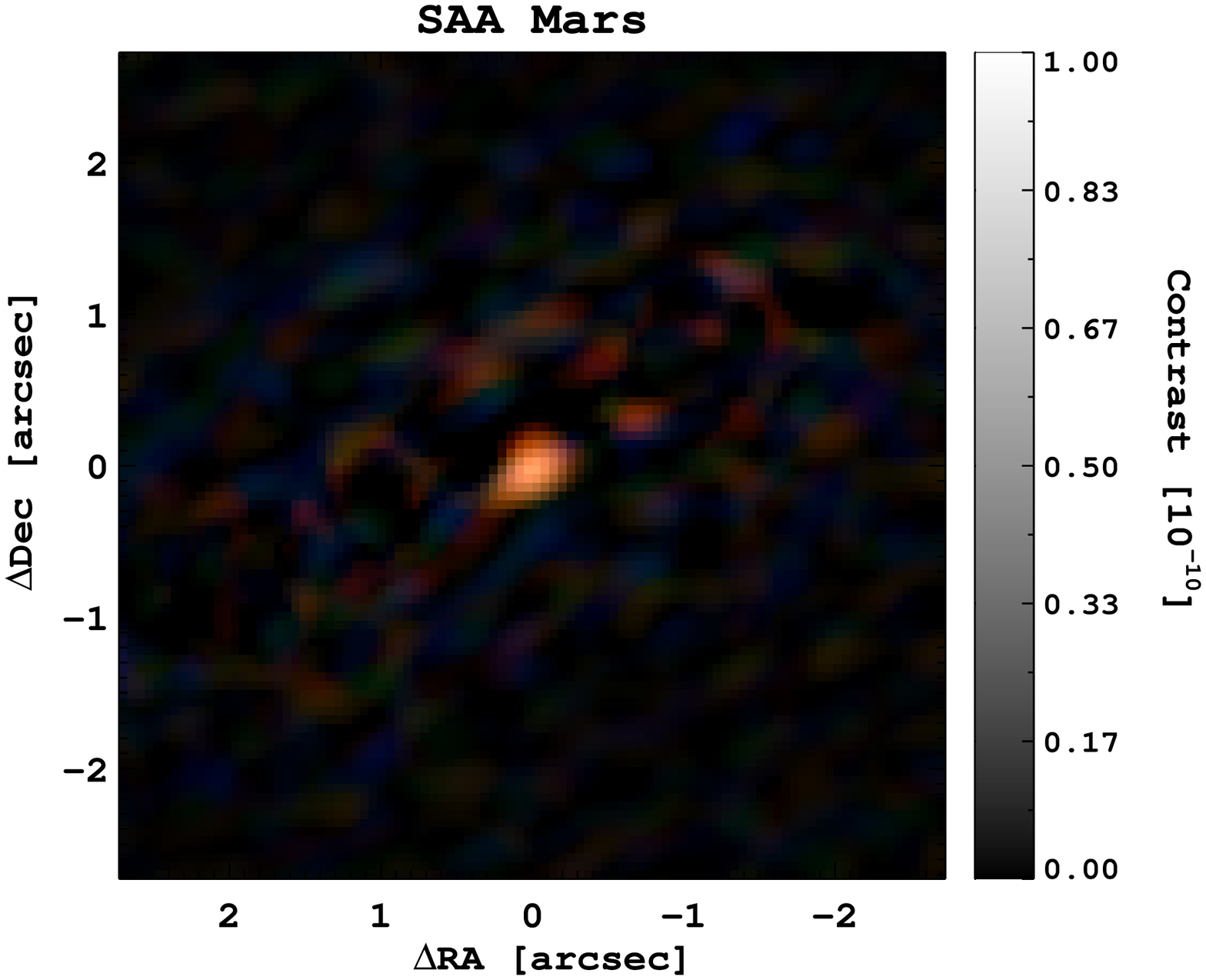}
   \caption[Simulation SAA] 
   { Images after de-orbiting and combining 2 years worth of data.  Each planet was coadded independently, with the frames being weighted by the phase function.  The pseudo-Mars planet is only marginally detected in any one frame, but is easily seen in the combined image.  
     \label{fig:simulation_saa} 
   }
\end{figure}

\end{landscape}


\bibliography{males}   

\begin{thebibliography}{1}

\bibitem{bendek_spie15}
Bendek, E.~A., Belikov, R., Thoms, S.~J., Lozi, J., and Males, J.~R., ``{Space
  telescope design to directly image the habitable zone of Alpha Centauri},''
  {\em Proc. SPIE}~{\bf 9605} (2015).

\bibitem{belikov_spie15}
Belikov, R., Bendek, E.~A., Thoms, S.~J., Males, J.~R., and Lozi, J., ``{How to
  directly image a habitable planet around Alpha Centauri with a $\sim$30-45cm
  space telescope},'' {\em Proc. SPIE}~{\bf 9605} (2015).

\bibitem{2006ApJ...641..556M}
{Marois}, C., {Lafreni{\`e}re}, D., {Doyon}, R., {Macintosh}, B., and {Nadeau},
  D., ``{Angular Differential Imaging: A Powerful High-Contrast Imaging
  Technique},'' {\em ApJ}~{\bf 641},  556--564 (Apr. 2006).

\bibitem{2013ApJ...771...10M}
{Males}, J.~R., {Skemer}, A.~J., and {Close}, L.~M., ``{Direct Imaging in the
  Habitable Zone and the Problem of Orbital Motion},'' {\em ApJ}~{\bf 771},  10
  (July 2013).

\bibitem{2012ApJ...755L..28S}
{Soummer}, R., {Pueyo}, L., and {Larkin}, J., ``{Detection and Characterization
  of Exoplanets and Disks Using Projections on Karhunen-Lo{\`e}ve
  Eigenimages},'' {\em ApJ}~{\bf 755},  L28 (Aug. 2012).

\bibitem{2010MNRAS.405.1907B}
{Bruntt}, H., {Bedding}, T.~R., {Quirion}, P.-O., {Lo Curto}, G., {Carrier},
  F., {Smalley}, B., {Dall}, T.~H., {Arentoft}, T., {Bazot}, M., and {Butler},
  R.~P., ``{Accurate fundamental parameters for 23 bright solar-type stars},''
  {\em MNRAS}~{\bf 405},  1907--1923 (July 2010).

\bibitem{2002A&A...386..280P}
{Pourbaix}, D., {Nidever}, D., {McCarthy}, C., {Butler}, R.~P., {Tinney},
  C.~G., {Marcy}, G.~W., {Jones}, H.~R.~A., {Penny}, A.~J., {Carter}, B.~D.,
  {Bouchy}, F., {Pepe}, F., {Hearnshaw}, J.~B., {Skuljan}, J., {Ramm}, D., and
  {Kent}, D., ``{Constraining the difference in convective blueshift between
  the components of alpha Centauri with precise radial velocities},'' {\em
  A\&A}~{\bf 386},  280--285 (Apr. 2002).

\bibitem{2014ApJ...786...32M}
{Males}, J.~R., {Close}, L.~M., {Morzinski}, K.~M., {Wahhaj}, Z., {Liu}, M.~C.,
  {Skemer}, A.~J., {Kopon}, D., {Follette}, K.~B., {Puglisi}, A., {Esposito},
  S., {Riccardi}, A., {Pinna}, E., {Xompero}, M., {Briguglio}, R., {Biller},
  B.~A., {Nielsen}, E.~L., {Hinz}, P.~M., {Rodigas}, T.~J., {Hayward}, T.~L.,
  {Chun}, M., {Ftaclas}, C., {Toomey}, D.~W., and {Wu}, Y.-L., ``{Magellan
  Adaptive Optics First-light Observations of the Exoplanet {$\beta$} Pic B. I.
  Direct Imaging in the Far-red Optical with MagAO+VisAO and in the Near-ir
  with NICI},'' {\em ApJ}~{\bf 786},  32 (May 2014).

\end{thebibliography}
\bibliographystyle{spiebib}   

\end{document}